\documentclass[a4paper, amsfonts, amssymb, amsmath, reprint, showkeys, nofootinbib, twoside, floatfix]{revtex4-1}

\usepackage[english]{babel}
\usepackage[utf8]{inputenc}
\usepackage[colorinlistoftodos, color=green!40, prependcaption]{todonotes}
\usepackage{amsmath}
\usepackage{mathtools}
\usepackage{xcolor}
\usepackage{graphicx}
\usepackage{subfigure}
\usepackage[left=23mm,right=13mm,top=35mm,columnsep=15pt]{geometry} 
\usepackage{adjustbox}
\usepackage{placeins}
\usepackage[T1]{fontenc}
\usepackage{lipsum}
\usepackage{csquotes}
\usepackage[pdftex, pdftitle={Article}, pdfauthor={Author}]{hyperref} 
\begin{document}
\title{In Search of Global 21-cm Signal using Artificial Neural Network in light of ARCADE 2}

\author{Vivekanand Mohapatra}
    \email[Correspondence email address: ]{p22ph003@nitm.ac.in}
    \affiliation{Department of Physics, National Institute of Technology Meghalaya, Shillong, Meghalaya, India}

\author{Johnny J}
    \affiliation{Jodrell Bank Centre for Astrophysics, Department of Physics \& Astronomy, University of Manchester, Oxford Road, Manchester M13 9PL, UK}

\author{Pravin Kumar Natwariya}
    \affiliation{School of Fundamental Physics and Mathematical Sciences, Hangzhou Institute for Advanced Study, UCAS, Hangzhou, China}
    \affiliation{University of Chinese Academy of Sciences, Beijing, China}

\author{Jishnu Goswami}
    \affiliation{RIKEN Center for Computational Science, 7-1-26 Minatojima-minamimachi,\\ Chuo-ku, Kobe, Hyogo 650 0047, Japan}

\author{Alekha C. Nayak}
    \email[Correspondence email address: ]{alekhanayak@nitm.ac.in}
    \affiliation{Department of Physics, National Institute of Technology Meghalaya, Shillong, Meghalaya, India}

\date{\today} 

\begin{abstract}
Understanding the astrophysical nature of the first stars remains an unsolved problem in cosmology. The redshifted global 21-cm signal $({T}_{21})$ acts as a treasure trove to probe the cosmic dawn era--- when the intergalactic medium was mostly neutral. Many experiments, like SARAS 3, EDGES, and DARE, have been proposed to probe the cosmic dawn era. However, extracting the faint cosmological signal buried inside a brighter foreground, $\mathcal{O}(10^4)$, remains challenging. Additionally, an accurate modelling of foreground and ${T}_{21}$ signal remains the heart of any extraction technique. In this work, we constructed the foreground signal $(T_{FG})$ from the global sky model and star formation history using Press-Schechter formalism to determine the $T_{21}$ signal with excess radio background following ARCADE 2 detection. Further, we incorporated static ionospheric distortion into the total signal and calculated the signal measured by an ideal antenna. We then trained an artificial neural network (ANN) for the extraction of a $T_{21}$ signal parameters signal measured by antenna with an R-square score $(0.5523 - 0.9901)$. Lastly, we used a Bayesian technique to extract $T_{21}$ signal and compared the finding with ANN's extraction.
\end{abstract}

\keywords{ cosmology: first stars, observations, cosmological parameters, method: data analysis}

\maketitle

\section{Introduction}\label{Intro}

The formation and evolution of the first astrophysical object in the universe remains an outstanding problem. Due to the uncertainties in the known physics of the formation, the thermal and ionization evolution of the intergalactic medium (IGM) during the cosmic dawn (CD) era and epoch of reionization (EoR) lacks a comprehensive understanding. Detection of the global 21-cm signal from these eras can shed light on it \cite{Furlanetto:2006wp, Pritchard:2011xb}. Many experiments have been conducted, for example, SARAS \cite{Patra:2012yw}, SCI-HI \cite{Voytek:2013nua}, SARAS 2 \cite{Singh:2017gtp}, HERA \cite{DeBoer:2016tnn}, SARAS 3 \cite{Singh:2021mxo}, REACH \cite{deLeraAcedo:2022kiu} to measure the global 21-cm signal $(\text{T}_{21})$. Recently, EDGES collaboration has reported a 21-cm signal with an absorption amplitude $\sim 0.5^{-0.2}_{+0.5}$ K in the redshift range $15-20$--- which is twice the value predicted by the standard $\Lambda$CDM model of cosmology \cite{Bowman:2018yin}. Although SARAS 3 have rejected the EDGES signal with a $95.3\%$ confidence level after conducting an independent check \cite{Bevins:2022ajf}, the actual shape is still unknown. If future experiments confirm a trough of amplitude greater than $\sim0.2$ K, which is the standard value \cite{Pritchard:2011xb}, that could lead to a completely new insight into the physics of these epochs. Following the EDGES detection, several models have been proposed to construct all possible amplitudes of the global 21-cm signal to probe exotic physics. For instance, to explain the EDGES trough, authors in Ref.  \cite{Bhatt:2019lwt} have proposed DM-Baryon Scattering in the presence of primordial magnetic fields. In contrast, authors in Ref. \cite{Barkana:2018qrx, Slatyer:2018aqg} have considered excess cooling of IGM by DM-Baryon scattering. Along with these IGM-cooling effects, energy injection by Primordial Black Holes \cite{Natwariya:2021xki, Saha:2021pqf} and decaying sterile neutrinos \cite{Natwariya:2022xlv} have also been considered to explain the EDGES trough. 

In addition to that, the presence of excess-radio background radiation (EBR) \cite{Fialkov:2018xre, Fraser:2018acy, Feng:2018rje, Pospelov:2018kdh} could also explain the EDGES trough. The existence of EBR could be due emission of radio photons from the conversion of axion-photon in the presence of magnetic field \cite{Moroi:2018vci}, Bose (axion) star \cite{Levkov:2020txo}, Pop III black holes \cite{Mebane:2020jwl}, and accreting PBHs \cite{Mittal:2021dpe}. Recently, ARCADE 2 \cite{Fixsen:2009xn, Feng:2018rje} and LWA1 \cite{Dowell:2018mdb} have detected an excess-radio background in the frequency range of $3-10$ GHz and $40-80$ MHz, respectively. ARCADE 2 detection mimics the Cosmic Microwave Background radiation (CMB) for a frequency $\nu>10$ GHz but deviated significantly otherwise. A power law has modelled these detections with a spectral index $(\beta)$ of $-2.62\pm0.04$ and $-2.58\pm0.05$ for ARCADE 2 \cite{Feng:2018rje} and LWA1 \cite{Dowell:2018mdb}, respectively.

Detecting the global 21-cm signal (mK) buried in a sea full of bright foreground radiation of the $\mathcal{O}(10^4)$ stronger is an observational challenge. In addition to that, the ionospheric effect and radio frequency interference (RFI) make it a tougher job. These effects can be reduced significantly if one considers farside Moon-based experiments like DARE \cite{Burns:2011wf, Burns:2017ndd}. One of the most common techniques adapted to remove the foreground radiation is considering it to be well-characterized and spectrally smooth. After removing the foreground radiation, the residual contains the global 21-cm signal having the signature of the IGM evolution. Machine Learning (ML) techniques have been used previously by many authors to study the CD and EoR \cite{Shimabukuro:2017jdh, Schmit:2017pho, Jennings:2018eko, Hassan:2018bbm, Chardin:2019euc, Gillet:2018fgb}. Artificial neural networks (ANN) are used for parameter extraction from the 21-cm power spectrum \cite{Shimabukuro:2017jdh, Schmit:2017pho}. Convolutional neural network (CNN) has also been used to study, emulate, and extract parameters from the 21-cm maps \cite{Jennings:2018eko, Hassan:2018bbm, Chardin:2019euc, Gillet:2018fgb}. 
To generate fast and accurate realizations of global 21-cm signal, machine learning-based emulators have been constructed, for example, 21cmGEM \cite{10.1093/mnras/staa1530}, GLOBALEMU \cite{Bevins:2021eah}, and 21cmVAE \cite{Bye:2021ngm}. Provided seven astrophysical free parameters, 21cmGEM \cite{10.1093/mnras/staa1530} is capable of producing global 21-cm signals in redshift $\text{z} = 5-50$ using a series of neural networks, a tree classifier, and principal component analysis. The objective of these emulators is to learn the relationship between the parameters and the global 21-cm signal without enforcing a physical model. However, it should be noted that these emulators can establish a relationship between $\text{T}_{21}$ signal and the parameters based on the signals they are trained with. Therefore, any possible scenario that can affect the $\text{T}_{21}$ signal significantly could also alter the parameters and their relationship with the signal as well.

This work focuses on extracting a global 21-cm signal from the CD era in the presence of contamination, such as foregrounds and ionospheric distortion. We first consider different realisations of $T_{21}$ signal representing various star formation histories in the presence of excess radio background following ARCADE 2 detection. We then model foreground signals from the global sky model and considered a static and homogeneous ionosphere to incorporate an additional distortion. The convolution of the foreground and 21-cm signal with ionospheric distortion defines the antenna temperature, which is observed by ground-based telescopes. In order to extract 21-cm signals without knowing the true nature of the foreground, we then train an ANN with antenna temperatures to estimate the associated 21-cm signal parameters. Additionally, we extracted the same 21-cm signal from the antenna temperature by deploying Markov Chain Monte Carlo (MCMC) technique. We find that contrary to ANN, MCMC can provide posterior distributions associated with the free parameters with uncertainties while being comparatively computationally expensive. On the other hand, unlike MCMC, ANN does not require a defined mathematical function to establish a relationship between antenna temperature and 21-cm signal parameters. Therefore, we can conclude that both techniques can complement each other in the extraction of a global 21-cm signal buried inside a strong foreground.

This work is organised as follows: Section \eqref{The Global 21cm Signal} briefly introduces the global 21-cm signal and excess background radiation from ARCADE 2. Section \eqref{Evolution of Gas Temperature}, the evolution of the IGM temperature in the presence of star formation. Section \eqref{Generating the global 21-cm Signal}-\eqref{Foreground modelling} represents the construction of the global 21-cm and foreground signals in the presence of ionospheric distortion and defined antenna temperature. In section \eqref{MCMC_section}, we deploy Markov chain Monte Carlo analysis on the antenna temperature to extract the foreground and 21-cm signal signal. In section \eqref{ANN}, we briefly introduce the ANNs, construction of training and prediction data sets. In section \eqref{Result}, we estimate the $T_{21}$ parameters from the antenna temperature, which includes ionospheric distortion. Section \eqref{summary} includes a summary, conclusion and outlook.

\section{The global 21-cm signal}\label{The Global 21cm Signal}

The hyperfine transition between singlet\,$(F=0)$\, and triplet\,$(F=1)$\, states in a neutral hydrogen atom (HI) occur due to the interaction of proton and electron spin. The relative number density of neutral hydrogen in triplet $(n_1)$ and singlet $(n_0)$ states is characterized by spin temperature\,$(T_s)$\,
\begin{equation}
    \frac{n_1}{n_0}= \frac{g_1}{g_0}e^{-2\pi\nu_{21}/T_s} = 3\times e^{-T_{*}/T_s}
\end{equation}
where, $g_1=3$\,, $g_0=1$\,denotes statistical weight of the respective states,\, $\nu_{21}=1420$\,MHz is frequency of the photon, and \,$T_{*} = 2\pi\nu_{21}$\,. The global 21-cm signal $(\text{T}_{21})$, measured relative to the cosmic background radiation \cite{Furlanetto:2006fs,Mesinger:2007pd, Mesinger:2010ne,  Pritchard:2011xb} is given by:

\begin{equation}
    \text{T}_{21}\approx 27 x_{\text{HI}} \left(\frac{0.15}{\Omega_mh^2}\frac{1+z}{10}\right)^{1/2}\left(\frac{\Omega_bh^2}{0.023}\right)\left(1-\frac{T_r}{T_s}\right)\text{mK}
    \label{deltaTb}
\end{equation}
where, $\Omega_b = 0.04897$ and $\Omega_m = 0.30966$ represent baryon and total matter density parameter in the unit of critical density, $h = 0.6766$  represents the Hubble parameter \cite{Planck:2018vyg}, $x_{\text{HI}}$\, denotes the fraction of neutral hydrogen atom, and $T_r$ denotes background radiation. After recombination, the baryon number density predominantly contained neutral hydrogen atoms thus making the global 21-cm signal a useful probe for studying dark ages, cosmic dawn, and EoR. It can provide information about first-star formation, X-ray and Ly$\alpha$ heating, radio background heating, and other exotic heatings. Throughout this work, we use Eq. \eqref{deltaTb} to construct the global 21-cm signal.

In the review \cite{Pritchard:2011xb}, the evolution of the global 21-cm signal in CMB bath has been described in detail; here, we will briefly explain the same. At the end of recombination, which occurred at a redshift\,$(z\approx 1100)$,\, neutral hydrogen atoms were formed, and the photons were free to travel, referred to as CMB. This period is often called the last scattering surface. Due to the efficient Compton scattering, the CMB and IGM were in thermal equilibrium, $T_{\text{CMB}}\approx T_{g}$, till $z\sim 300$ causing an absence of the 21-cm signal $\text{T}_{21}$. Due to cosmic expansion, $T_{g}$ and $T_{\text{CMB}}$ cooled down $\propto(1+z)^2$ and $(1+z)$ respectively, over the time. For redshifts $z>100$, the collisional coupling $x_c\gg 1$ is thought to have caused an absorption signal. Nevertheless, this absorption signal has not been observed yet due to radio antennas' poor sensitivity, which falls dramatically for frequencies below \,$50\,\rm MHz$. At $z<40$ till the formation of the first star, the $x_c\sim 0$ causes no 21-cm signal \cite{Pritchard:2011xb, Barkana:2018qrx}. Ly$\alpha$ radiation coupled the gas via the WF effect after the first star formation at redshift \,$z\sim 30-25$\, making\, $x_{\alpha}\gg 1$. During this time, an absorption signal can be observed, and this phase is called the Cosmic Dawn (CD). Around \, $z\sim 15$, X-ray radiation from AGN could have heated the gas, causing an emission signal. After a certain period, at\, $z\sim 7-5$\,; $x_e\sim 1$\, resulting in no signal. This era is called the Epoch of Reionization (EoR). Instead of considering a wide range of redshifts, in this work, we consider EDGES reference \, $z\sim 27-14$ \cite{Bowman:2018yin}.

The most crucial quantities in Eq. \eqref{deltaTb} are $x_{\text{HI}}$, $T_s$, and $T_r$ which determines the intensity of $\text{T}_{21}$. The spin temperature $(T_s)$ given by,

\begin{equation}
    T_s^{-1} = \frac{T_r^{-1}+x_{\alpha}T_{\alpha}^{-1}+x_cT_g^{-1}}{1+x_{\alpha}+x_c},
    \label{spin_temp}
\end{equation}
where, $T_{\alpha}$ and $T_g$ respectively represent the colour temperature of Ly$\alpha$ radiation field and kinetic temperatures of IGM. Typically, $T_{\alpha}\approx T_g$, because the optical depth of Ly$\alpha$ photons is large, which leads to a large number of scattering, thus bringing the radiation field and IGM to local equilibrium \cite{Furlanetto:2006fs, Pritchard:2005an}. In this work, we use this approximation. The terms $x_{\alpha}$ and $x_c$ denote the Ly$\alpha$ and collisional coupling coefficient about the excess-radio background radiation \cite{Hirata:2005mz, Mesinger:2010ne} respectively. The collisional coupling coefficient is expressed as,

\begin{equation}
    x_{c} = \frac{T_{*}}{T_r}\frac{n_ik_{10}^{iH}}{A_{10}},
    \label{eq:xc}
\end{equation}
where, $n_i$, $k_{10}^{iH}$, and $A_{10}= 2.85\times 10^{-15} ~\rm Hz$ represent the number density of the species `$i$', the spin de-excitation rate coefficient due to collisions of species `$i$' with the hydrogen atom, and the Einstein coefficient for spontaneous emission from triplet to the singlet state, respectively. Calculating the de-excitation rate requires quantum mechanical calculations, whose tabulated values corresponding to $k^{HH}_{10}$ and $k^{eH}_{10}$ can be found in literature \cite{2006nla..conf..296Z, 2007MNRAS.374..547F, 2007MNRAS.379..130F}. However, in this work, we use an approximated functional form of $k^{HH}_{10}$ and $k^{eH}_{10}$, which are expressed as \cite{2006ApJ...637L...1K, 2001A&A...371..698L, Pritchard:2011xb},

\begin{alignat}{2}
	k^{HH}_{10} & = 3.1 \times 10^{-17}\left(\frac{T_{g}}{\mathrm{K}}\right)^{0.357}\cdot e^{-32\mathrm{K} / T_g}, \\
	\log_{10}{k^{eH}_{10}} & = -15.607 + \frac{1}{2}\log_{10}\left(\frac{T_g}{\mathrm{K}}\right)\times \nonumber \\ & ~~~~\qquad\qquad \exp{-\dfrac{\left[\log_{10} \left(T_g/\mathrm{K}\right)\right]^{4.5}}{1800}}.
    \label{coupling_coefficients}
\end{alignat}
Here, all $k^{iH}_{10}$ terms have the dimension of $\rm m^3 \rm s^{-1}$. We can now rewrite Eq. \eqref{eq:xc} as

\begin{equation}
	x_{c} = \frac{n_HT_{*}}{A_{10}T_{\gamma}}\left\{(1 - x_e)k_{10}^{HH}+ x_e\,k^{eH}_{10} \right\}, 
\end{equation}
where $n_H$ represents the number density of hydrogen atoms.

During the cosmic dawn era, $T_s$ is primarily affected by the Ly$\alpha$ photon field from astrophysical sources via the Wouthuysen-Field (WF) effect \cite{1952AJ.....57R..31W,1959ApJ...129..536F}. Therefore, to evaluate $x_c$ we closely follow Ref. \cite{Mittal:2020kjs}. The Ly$\alpha$ coupling can be expressed as \cite{Hirata:2005mz, Mesinger:2010ne, Pritchard:2011xb},

\begin{equation}
    x_{\alpha} = \frac{T_{*}}{T_r}\frac{4P_{\alpha}}{27 A_{10}},
\end{equation}
where $P_{\alpha}$ represents the total rate of Ly$\alpha$ photon scattering
per hydrogen atom. $P_{\alpha}$ depends on specific intensity $(J_{\alpha})$ of the Lyman alpha photons, and for that, we need to calculate the emissivity $\epsilon_{\alpha}$ of the Ly$\alpha$ photons \cite{Barkana:2004vb}. To begin with, we first consider population II stars and their spectral energy distribution as $\phi(\alpha) = 2902.91 \Tilde{E}^{-0.86}$ \cite{Mittal:2020kjs}. The terms $\Tilde{E} = E/E_{\rm ion}$ and $E\in [E_{\alpha}, E_{\beta}]$, where, $E_{\alpha} = 10.2\,\rm eV$, $E_{\beta} = 12.09\,\rm eV$, and $E_{\rm ion} = 13.6\,\rm eV$ represent energies of Ly$\alpha$ photon, Ly$\beta$ photon, and Lyman limit transition, respectively. We can now define $\epsilon_{\alpha}$ as \cite{Mittal:2020kjs}

\begin{equation}
    \epsilon_{\alpha}(E, z) = f_{\alpha}\phi_{\alpha}(E) \dot{\rho}_{*}(z)/m_b,
\end{equation}
where $m_b$, $f_{\alpha}$, and $\dot{\rho}_{*}$ represent baryon's mass, scaling parameter for $\phi_{\alpha}$, and star formation rate density (SFRD), respectively. The SFRD is determined by the rate at which baryons collapse into dark matter haloes \cite{Barkana:2004vb}. The number of haloes at redshift $z$ can be determined by the Press-Schechter formalism \cite{Press:1973iz}. The SFRD can be expressed as

\begin{equation}
    \dot{\rho}_{*}(z) = -f_{*} \bar{\rho}_{b}^{0}(1 + z)H(z) \frac{dF_{\rm coll}(z)}{dz},
\end{equation}
where $\bar{\rho}_{b}^{0} = \rho_c\Omega_{b,0}$ and $\rho_c$ represent baryon and critical density today, respectively. The term $f_{*}$ represents star formation efficiency. The fraction of baryons that have collapsed into dark matter haloes $(F_{\rm coll})$ is given by \cite{Barkana:2000fd}

\begin{equation}
    F_{\rm coll}(z) = \rm erfc \left[\frac{\delta_c(z)}{\sqrt{2}\sigma(m_{\rm min})}\right],
    \label{eq:F_coll}
\end{equation}
where $\delta_c$, $\sigma^2$, and $\rm erfc(\cdot)$ represent the linear critical density of collapse, the variance in the smoothed density field, and the complementary error function, respectively. The virial temperature $(\rm T_{\rm vir})$ of dark matter haloes is incorporated into the model through the minimum halo mass for star formation, which can be expressed as \cite{Mittal:2021egv}

\begin{equation}
	\rm m_{\rm min} = \frac{10^8 \rm M_{\odot}}{\sqrt{\Omega_mh^2}}\left[\frac{10}{1+z}\frac{0.6}{\mu} \frac{\rm min(\rm T_{\rm vir})}{1.98\times 10^4}\right]^{3/2}\, ,
\end{equation}

where $\rm M_{\odot}$ represents the solar mass and $\mu\approx 1.22$ \cite{DAYAL20181}. The term $\rm min(T_{\rm vir}) = 10^4\, \rm K$ represents the minimum virial temperature of dark matter haloes hosting star formation. In this work, we consider haloes with virial temperature $\rm T_{\rm vir}\geq 10^4\,\rm K$. To calculate $\delta_{c}/\sigma(m_{\rm min})$ we use the \texttt{COLOSSUS} software \cite{Diemer:2017bwl}. After defining SFRD, we can now evaluate $J_{\alpha}$. The Ly$\alpha$ specific intensity can be defined as \cite{Hirata:2005mz, Pritchard:2011xb},

\begin{equation}
    J_{\alpha} = \frac{c}{4\pi}(1+z)^2\sum_{n = 2}^{23}P_n\int_{z}^{z_{\rm max}}\frac{\epsilon_{\alpha}(E_n',z')}{H(z')}\, dz,
    \label{eq:J_alpha}
\end{equation}
where $P_n$ represents a finite probability at which an upper Lyman series photon redshifts to Ly$\alpha$ wavelength before getting absorbed or scattered. The values of $P_n$ are tabulated in articles \cite{Hirata:2005mz, Pritchard:2005an}. 
The redshifting energy of a photon $(E_n')$ originated at redshift $z'$ will have an energy, $E_n$, at redshift $z$. This can be expressed as $E_n' = E_n(1+z')/(1+z)$, where, $E_n$ represents photon's energy transiting from $n^{\rm th}$ to the ground state of a hydrogen atom. Furthermore, the upper limit of the integral in Eq. \eqref{eq:J_alpha} can be evaluated as \cite{Mittal:2020kjs}

\begin{equation}
    1+z_{\rm max} = \frac{E_{n+1}}{E_n}(1+z) = \frac{1-(1+n)^{-2}}{1- n^{-2}}(1+z).
\end{equation}

The Ly$\alpha$ coupling coefficient can be rewritten as (for a detailed discussion, refer \cite{Mittal:2020kjs}) 

\begin{equation}
    x_{\alpha} = \frac{SJ_{\alpha}}{J_0},
\end{equation}
where $J_0 \approx 5.54\times 10^{-8} T_r/T_{\rm CMB}\,\rm m^{-2}s^{-1}Hz^{-1}sr^{-1}$, and $S$ is called scattering correction. In this work, we take $S\sim 1$. Here, $T_{\text{CMB}} = T_{\text{CMB},0}(1+z)$ represents cosmic microwave background radiation and $T_{\text{CMB},0} = 2.725\,\rm K$. Below, we will discuss the possibilities of excess radio background radiation apart from CMB.

Earlier in this section, we have defined $T_r$ in Eq. \eqref{spin_temp} as the background radiation. In a standard scenario $T_r = T_{\text{CMB}}$, however, the existence of excess background radiation (EBR) cannot be denied completely. In Ref.\cite{10.1093/mnras/sty016}, authors have modelled EBR assuming local star formation rate at frequency $150\,\rm MHz$, whereas ARCADE 2 \cite{Fixsen:2009xn, Feng:2018rje} and LWA1 \cite{Dowell:2018mdb} measurements inspire a uniform redshift-independent synchrotron-like radiation whose phenomenological model can be given by \cite{Yang:2018gjd, Fialkov:2019vnb, Mondal:2020rce, Reis:2020arr,  Banet:2020ele}

\begin{equation}
    T_r = T_{\text{CMB},0}+ T_0\left(\frac{\nu}{\nu_0}\right)^{\beta},
    \label{bg_radiation}
\end{equation}
where $T_0 = 24.1\,\rm K$, the reference frequency $(\nu_0 = 310\,\rm MHz)$, and $\beta=-2.6$ is the spectral index \cite{Dowell:2018mdb}. For redshifted 21-cm photons, the observed frequency $\nu = 1420\text{ MHz}/(1+z)$, thus, we can rewrite the above equation as
\begin{equation}
    \frac{T_r}{T_{\text{CMB}}} = \left[1+0.169\, \zeta_{\rm ERB}\left(1+z\right)^{2.6}\right],
    \label{mod_bg_radiation}
\end{equation}
We note that for $\zeta_{\rm ERB} = 1$, the excess radiation satisfies ARCADE 2 measurement \cite{Fixsen:2009xn}.

\section {Evolution of Gas Temperature}\label{Evolution of Gas Temperature}

To compute the global 21-cm signal we need $T_g(z)$ and $x_e(z)$.
The evolution of gas temperature $(T_g)$ is expressed as

\begin{alignat}{2}
    \frac{dT_g}{dz} = 2\frac{T_g}{1+z} &+ \frac{\Gamma_c}{(1+z)H} \left(T_g - T_{\text{CMB}}\right)\nonumber \\ &+ \frac{2}{3n_bk_B(1+z)H}\sum q,
    \label{Gas Evolution}
\end{alignat}
where the first and second terms on the right-hand side represent the adiabatic cooling of the gas and the coupling between CMB and IGM due to Compton scattering, respectively \cite{Seager:1999bc, Pritchard:2011xb}. The third term accounts for all other heating and cooling processes. In this work, we consider X-ray heating from astrophysical sources. The Compton scattering rate $(\Gamma_c)$ is defined as,
\begin{equation*}  
    \Gamma_c = \frac{8 n_e\sigma_T a_rT_{\gamma}^4 (z)}{3m_e n_{\text{tot}}} ,\label{Compton_scattering}
\end{equation*}
where $m_e$, $\sigma_T$, and $f_{He} = 0.08$ represent the rest mass of an electron, Thomson scattering cross-section, and helium fraction, respectively. Whereas, $a_r = 7.57\times 10^{-16}$~J\,$\text{m}^{-3}\,\text{K}^{-4}$ represents the radiation density constant and $N_{\rm tot} = N_H(1+f_{He}+x_e)$ represents the total number density of gas \cite{Seager:1999bc, Seager:1999km}. The evolution of the ionization fraction can be expressed as \cite{Peebles:1968ja, Seager:1999bc},

\begin{equation}
	\frac{dx_e}{dz} = \frac{\mathcal{P}}{(1+z)H} \left[n_Hx_e^2\alpha_B - (1-x_e)\beta_Be^{-E_{\alpha}/T_{\gamma}}\right],
	\label{xe_evolution}
\end{equation}

here $\mathcal{P}$ represents Peebles coefficient, while $\alpha_B$ and $\beta_B$ are the case-B recombination and photo-ionization rates, respectively \citep{Seager:1999bc, Seager:1999km, Mitridate:2018iag}. The Peebles coefficient is given by \cite{Peebles:1968ja, DAmico:2018sxd},
\begin{equation*}
    \mathcal{P} = \frac{1+ \mathcal{K}_H\Lambda_Hn_H(1-x_e)}{1+ \mathcal{K}_H(\Lambda_H+\beta_H)n_H(1-x_e)},
\label{peeble_coefficient}
\end{equation*}
where $\mathcal{K}_H = \pi^2/(E_{\alpha}^3H)$, $E_\alpha = 10.2\,\rm eV$, and $\Lambda_H = 8.22/\text{sec}$ represents redshifting Ly${\alpha}$ photons, rest frame energy of Ly$\alpha$ photon, and 2S-1S level two-photon decay rate in hydrogen atom respectively \cite{PhysRevA.30.1175}.
The influence of non-thermal excess-radio radiation $(T_r)$ on the gas temperature is insignificant (Eq. \ref{Gas Evolution}) and thus can be ignored \cite{Feng:2018rje}. 

Here, we discuss the implications of X-ray photons on the evolution of $T_g$ and $x_e$. X-ray photons have a large mean free path compared to Ly$\alpha$ photons. Therefore, these photons can travel far from the source to heat and partially ionize the IGM \cite{Mirabel:2011rx}. The astrophysical sources producing X-ray photons could be X-ray binaries and mini-quasars \cite{Madau:2003um, Power:2012hm, Fragos:2013bfa}. The X-ray heating mechanism can be described as follows: X-ray photons with a long mean free path can travel far away from their sources and photoionize the neutral hydrogen atoms. This produces energetic free electrons, which dissipate energy via atom excitations and collisions with the residual free electrons present in the IGM. In this process, the average kinetic energy of the IGM increases, which results in an increase in the IGM temperature. To relate the X-ray emissivity to the star formation rate (SFR), we assume that the SFR is directly proportional to the rate at which baryonic matter collapses onto virialized haloes, that is $dF_{\rm coll}/dt$ (Eq. \ref{eq:F_coll}). Following Ref. \cite{Furlanetto:2006jb}, We can now write the third term of Eq. \eqref{Gas Evolution} as

\begin{equation}
    \frac{2}{3}\frac{q_X}{k_Bn_b(1+z)H(z)} = 5\times 10^5\,\mathrm{K}\,(f_Xf_*f_{Xh}) \frac{dF_{\rm coll}}{dz},
    \label{X-ray_term_in_Tg}
\end{equation}
where $f_X$ is a normalization factor analogous to $f_{alpha}$ in case of Ly$\alpha$. $f_{X,h}$ represents the fraction of X-ray energy that goes into heating the IGM. We note that $f_X$ and $f_{X,h}$ are degenerated; therefore, we consider them together as a single parameter $f_xf_{Xh}$.

The next step is to evaluate the ionization fraction in the presence of X-ray sources. We consider that the ionizing photons are produced inside the galaxies, such that their production rate can be taken to be proportional to the star formation rate \cite{Furlanetto:2006jb}. The ionization efficiency $\xi_{\rm ion}$ can be expressed as 

\begin{equation}
    \xi_{\rm ion} = \mathrm{A_{He}}\rm f_*f_{\rm esc}N_{\rm ion},
\end{equation}
where, $\rm f_{\rm esc}$ is the fraction of ionizing photons escaping the host galaxies, $N_{\rm ion}$ is the number of ionizing photons produced per baryon, and $\mathrm{A_{He}} = 4/(4-3Y_p)$ is the mass fraction of helium. We can now rewrite the ionization fraction in the presence of X-rays as \cite{Furlanetto:2006jb}

\begin{equation}
    \frac{dx_e}{dz} = \frac{dx_e}{dz}\bigg{|}_{(\rm Eq. \ref{xe_evolution})}-\xi_{\rm ion} \frac{dF_{\rm coll}}{dz}.
    \label{xe_evolution_modified}
\end{equation}
As $f_{\rm esc}$ and $N_{\rm ion}$ values degenerated, therefore, we combined them $(f_{\rm esc}N_{\rm ion})$ to make consider them a single parameter and set it to unity. In the below section, we will evaluate $T_{21}$ in the presence of Ly$\alpha$ and X-ray radiations.

\section{21-cm signal modelling}\label{Generating the global 21-cm Signal}

\begin{figure*}[htbp]
    \centering
    \subfigure[]{
        \includegraphics[width=0.45\linewidth]{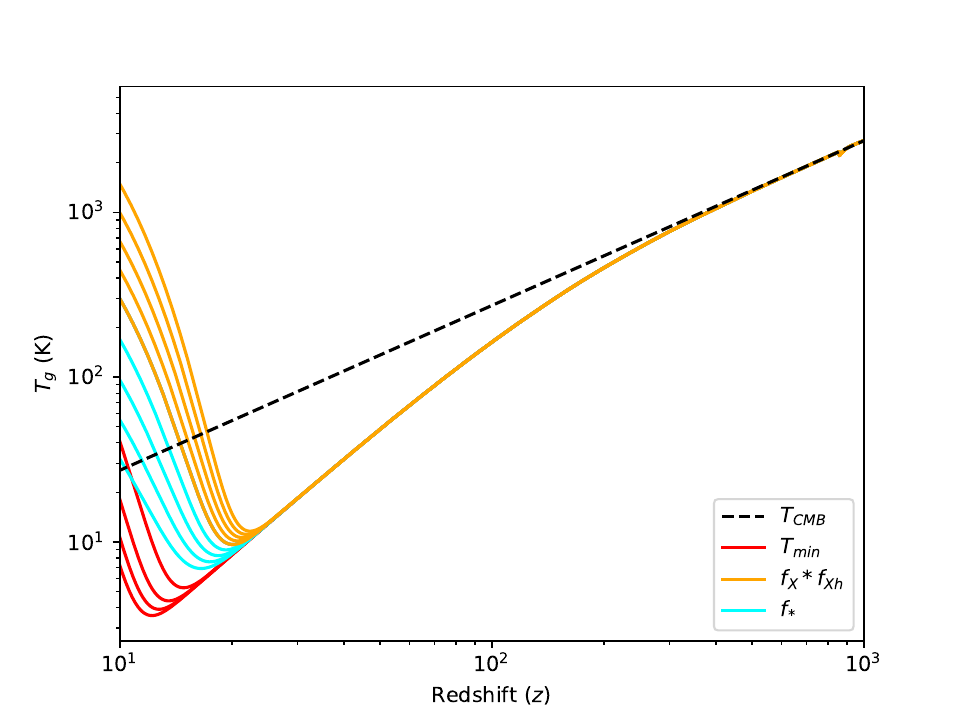}
        \label{fig:Tk_plots}
    }
    \hfill
    \subfigure[]{
        \includegraphics[width=0.45\linewidth]{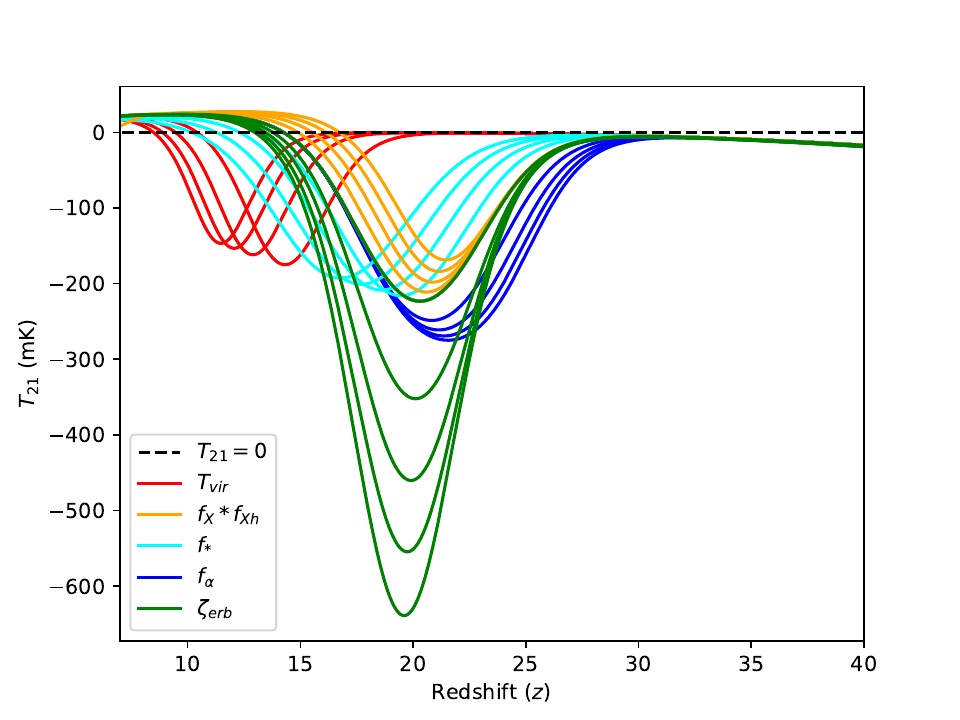}
        \label{fig:T21_plots}
    }
    
    \caption{(a) Thermal evolution of IGM temperature in the presence of X-ray heating. The red solid lines represent variation in $T_g$ for different $\rm T_{vir}$. The orange and solid lines represent different $\rm f_Xf_{Xh}$ and $\rm f_*$ values. The black dashed line is the CMB temperature. (b) This plot represents different $T_{21}$ signals during the cosmic dawn era. The red, orange, and cyan solid lines represent $T_{21}$ signal corresponding to the $T_g$ presented in the left panel. The blue and green solid lines represent variations in $f_{\alpha}$ and $\zeta_{\rm erb}$.} 
    \label{gas_evolution_plot}
\end{figure*}

Previous studies have incorporated different techniques to achieve different amplitudes of the global 21-cm signal. Authors in Ref. \cite{Fialkov:2013uwm, Fialkov:2015fua, Cohen:2016jbh} have predicted possible global 21-cm signals in the redshift $z \sim 6-40$ employing a semi-numerical technique. The signal is parameterized using physical characteristics--- closely connected to the IGM features, allowing us to deduce the physics of the earliest source. In addition, the absorption characteristic has also been represented as Gaussian \cite{Bernardi:2014caa, Bernardi:2016pva}. Similarly, a turning point model was proposed indicating positions and amplitude at redshifts and the shape of the 21-cm signal \cite{Pritchard:2010pa}. The section of the 21-cm  signal between the turning points is represented as a cubic spline, which allows great flexibility and may describe a variety of 21-cm signals. Nevertheless, additional interpretation is required to link the turning point positions to the physics of the initial luminous sources. Furthermore, authors in Ref. \cite{Mirocha:2013gvq, Mirocha:2015jra, Harker:2015oxa} have proposed a $\tanh$ parameterization which uses a succession of $\tanh$ functions to simulate the 21-cm signal. In this work, we consider an approach that is physically driven.

To evaluate the 21-cm global signal, we have to solve $T_g(z)$ and $x_e(z)$ simultaneously. We replaced the third term on the right-hand side of Eq. \eqref{Gas Evolution} with Eq. \eqref{X-ray_term_in_Tg} to include X-ray heating. We then solve the evolution the IGM temperature (Eq. \ref{Gas Evolution}) and ionization fraction (Eq. \ref{xe_evolution_modified}) with the initial conditions $T_g = 2967.6\text{ K}$, and $x_e = 0.1315$ at redshift $z = 1088$ taken from \texttt{Recfast++} \cite{Chluba:2010fy, Chluba:2010ca}. To monitor the variations in $T_g$ in the presence of different astrophysical scenarios, we varied $\rm T_{\rm vir}$, $f_*$, $\rm f_Xf_{Xh}$, $\rm f_{\alpha}$, and $\zeta_{\rm erb}$. In Fig. \eqref{fig:Tk_plots}, we plot the variations in $T_g$ for different values of the free parameters. For instance, we fix $\rm f_* = 0.1$, $\rm f_Xf_{Xh} = 0.2$ and vary $\rm T_{\rm vir}/10^4$ in range $1-50\,\rm K$--- shown red solid lines. Similarly, we fix $\rm T_{\rm vir} = 10^4\,\rm K$, $\rm f_Xf_{Xh} = 0.2$ and vary $\rm f_{*}$ in range $0.01 - 0.1$--- depicted in cyan solid lines. Lastly, we fix $\rm T_{\rm vir} = 10^4\,\rm K$, $\rm f_* = 0.1$, and vary $\rm f_Xf_{Xh}$ in range $0.2 - 1$--- presented in orange solid line. The black-dashed line represents the CMB temperature. We note that while plotting Fig.  \eqref{fig:Tk_plots} we keep $\rm f_{\alpha}$ fixed to unity. As we have not included heating due to Ly$\alpha$ photons, therefore, $T_g$ remains unaffected by $\rm f_{\alpha}$ values. Next, in Fig. \eqref{fig:T21_plots}, we show the variations in $T_{21}$ for different values of the aforementioned parameters. The red, cyan, and orange solid lines respectively represent $T_{21}$ signal corresponding to the $T_g$ shown in Fig. \eqref{fig:Tk_plots} for $\rm f_{\alpha} = 1$ and $\zeta_{\rm erb} = 10^{-3}$. We then fix $\rm T_{\rm vir}$, $\rm f_{*}$, $\rm f_Xf_{Xh}$, $\zeta_{\rm erb}$ to $10^4\,\rm K$, $0.1$, $0.2$, $10^{-3}$, respectively, and vary $f_{\alpha}$ in range $1-5$--- shown in blue solid lines. Finally, we fix $\rm T_{\rm vir}$, $\rm f_{*}$, $\rm f_Xf_{Xh}$, $\rm f_{\alpha}$ to $10^4\,\rm K$, $0.1$, $0.2$, $1$, respectively, and vary $\zeta_{\rm erb}$ in range $10^{-3}-10^{-2}$--- depicted in green solid lines. In the next section, we will model the foreground and noises from different sources.

\section{Foreground modelling} \label{Foreground modelling}

Detecting the weak global 21-cm signal during the CD era is challenging due to the presence of strong foreground radiation, instrumental systematics, radio frequency interference, and ionospheric distortions. The radio emissions from the Milky Way and other extragalactic sources are significantly brighter than the cosmological signal. The expected strength of the global 21-cm signal is approximately $10^{-4}$ times weaker than the foreground emissions. These challenges require sophisticated simulations to determine the impact of these factors on the extraction of the desired signal. Therefore, having a precise model of foregrounds at radio frequencies is crucial to ensure accurate extraction.
According to Ref. \cite{Pritchard:2010pa, Bernardi:2014caa, Bernardi:2016pva}, the foreground spectrum can be represented as a $\log$-polynomial between foreground temperature and frequency as $\ln(\text{T})-\ln(\nu)$. Authors in Ref. \cite{Harker:2015uma} have demonstrated that using a $3^{\text{rd}}$ or $4^{\text{th}}$-order polynomial is sufficient enough to model the sky spectrum. However, it has been found that a $7^{\text{th}}$-order polynomial is necessary when incorporating the chromatic primary beam of the antenna. SARAS 3 have used a $6^{\text{th}}$-order polynomial to model the galactic and extragalactic foreground passing through ionosphere added with systematic calibration error in a band of frequency $55-85$ MHz with a resolution of $61\,\rm kHz$ \cite{Singh:2021mxo}. Furthermore, a 5$^{\text{th}}$-order polynomial has also been used to represent the galactic synchronous radiation of spectral index $-2.5$ along with Earth's ionosphere distortion \cite{Bowman:2018yin}. Below, we will explain the foreground model considered in this work.

We consider the diffuse foregrounds to have the form of a polynomial in $\ln (T)-\ln (\nu)$, that is \cite{deOliveira-Costa:2008cxd, Harker:2015oxa, Datta:2014xla}

\begin{equation}
    \ln\,T_{FG} = \ln\,T_0 + \sum_{i = 1}^{3}a_i[\ln(\nu/\nu_0)]^{i},
    \label{eq:T_FG}
\end{equation}

where $\nu_0 = 80\,\rm MHz$ is an arbitrary reference frequency, and $\{T_0, a_1, a_2, a_3\}$ are the parameters of the model we will consider in this work. We note that on increasing the number of parameters, the foreground becomes more complicated and less smooth \cite{Harker:2015oxa}. We follow Ref. \cite{Harker:2015oxa} work to incorporate the four free parameters values, that is, $\{T_0, a_1, a_2, a_3\} = \{2039.611, -2.42096, -0.08062, 0.02898\}$. These values are computed by fitting the quiet region of the global sky model (GSM) convolved with a beam with a full width at half-maximum of $72^{\circ}$, with a third-order polynomial over $35-120\,\rm MHz$ \cite{deOliveira-Costa:2008cxd}. When the diffused radiations enter the Earth's ionosphere, they suffer refraction, absorption, and thermal emission due to the presence of energetic electrons, as well as electrically charged atoms and molecules. The ionosphere is heavily affected by the Sun's X-ray and Ultraviolet (UV) radiation, which increases the concentrations of free energetic electrons through photoionization. Therefore, the ionospheric effects on the total signal, that is, foreground plus $T_{21}$ signal, varies depending on solar activities (refer article \cite{Datta:2014xla} and references therein). For simplicity, we consider a static ionosphere in this work.

The ionosphere is primarily divided into F-layer and D-layer. Let us first discuss the F-layer, which can distort the incoming radio signals by refracting them. This layer is spanned over $\sim 200\,\rm Km$ to $\sim 400\,\rm Km$ above the Earth's surface. The refraction occurring at the ionosphere's F-layer can be analogous to the effect of a spherical lens, where a refracted ray is bent toward the zenith \cite{Vedantham:2013mea}. As a result of this refraction, ground-based radio antennas capture signals from a broader area of the sky, leading to an increase in the observed antenna temperature. Assuming the F-layer has a parabolic geometry and is bounded by free space with a refractive index of unity for radio waves, the angular deviation experienced by an incident ray at an angle $\theta$ relative to the horizon is given by \cite{Datta:2014xla}

\begin{equation}
    \delta\theta (\nu) = \frac{2d\cos\theta}{3\rm R_E} \left(\frac{\nu_p}{\nu}\right)^2 \left(1+\frac{h_m}{\rm R_E}\right)\left(\sin^2\theta + \frac{2h_m}{\rm R_E}\right)^{-3/2},
    \label{eq:delta_theta}
\end{equation}
here $d= 200\,\rm Km$ and $\rm R_E = 6378\,\rm Km$ denote the change in the altitude with respect to $h_m$ where the electron density goes to \textit{zero} and the Earth's radius, respectively. Whereas $h_m = 400\,\rm Km$ represents the height of the F-layer where the electron density is maximum. Here, $\nu$ represents the frequency of the incoming radio wave, and $\nu_p$ is the plasma frequency (per cubic meter) given by \cite{Datta:2014xla}

\begin{equation}
    \nu_p^2 (t) = \frac{e^2}{4\pi^2\epsilon_0m_e}n_{e,f}(t)\, ,
\end{equation}
where $e$, $m_e$, and $n_{e,f}$ represent the electron's charge, rest mass, and number density in the F-layer, respectively.
The electron density in the F-layer can vary over time; however, in this work, we have considered it to be static. From Eq. \eqref{eq:delta_theta}, it is evident that the maximum deviation occurs when the incident angle is $\theta= 0$, corresponding to the horizon ray. Now, for a given observational frequency, the field of view becomes greater than the antenna's primary beam due to the refraction. Thus, this effect adds a distortion to the incoming signal. Before discussing how the antenna temperature is affected, let us first discuss the effect of the D-layer on the incoming radio waves.

Due to the D-layer's lower altitude, it contains a higher concentration of neutral gas molecules, leading to an increased collision frequency $\nu_c$ of electrons. Since the electron density $n_{e,d}$ in the D-layer is closely linked to the neutral gas density, the following empirical formula for the collisional frequency $(\nu_c)$ of the electrons can be derived as \cite{Datta:2014xla}

\begin{equation}
    \nu_c = 3.64\left(\frac{n_{e,d}}{T_e^{3/2}}\right)\left\{19.8 + \ln \left(\frac{T_e^{3/2}}{\nu}\right)\right\}\,\rm Hz,
\end{equation}
where $T_e$ represents the absolute temperature of the D-layer
plasma, characterizing the thermal kinetic energy per particle. We have considered $n_{e,d} = 2.5\times 10^8$ and $T_e = 200\,\rm K$ as the typical values in the simplistic case. Consequently, the approximated refractive index of the D-layer as \cite{Vedantham:2013mea, Shen:2020jwt}

\begin{equation}
    \eta_D\approx -\frac{1}{2}\frac{(\nu_c/\nu)\nu_p^2}{\nu^2+\nu_c^2}.
\end{equation}
In a homogeneous ionospheric layer, the electric ﬁeld can be
described as a plane wave,

\begin{equation}
    E(\Delta s, \theta) = E_0\,\exp\left(-i\frac{2\pi\eta}{c}\Delta s (\theta)\right)
\end{equation}

where, $c$ represents the speed of light in free space, $\Delta s$ is the path length along the wave's propagation direction within the ionosphere, and $E_0$ denotes the initial electric field when $\Delta s = 0$. Since $\eta$ is imaginary, it leads to exponential attenuation of the wave, resulting in absorption \cite{Vedantham:2013mea}. Given that the intensity of an electromagnetic wave is proportional to the square of its amplitude, we define the loss factor $\mathcal{L}$ as the remaining portion of the wave after being absorbed by the ionosphere. The loss factor is given by

\begin{equation}
    \mathcal{L}(\nu,\theta) = \exp\left[\frac{4\pi\nu\eta_D}{c}\Delta s(\theta)\right].
\end{equation}
The length of the path traversed by an electromagnetic wave propagating through the ionosphere at a given incident angle $\theta$ can be approximated using an expression derived from the model's geometric framework as \cite{Vedantham:2013mea}

\begin{equation}
    \Delta s(\theta)\approx \Delta h_D\left(1+\frac{h_D}{\rm R_E}\right)\left(\cos^2\theta+\frac{2h}{\rm R_E}\right)^{-1/2},
\end{equation}
where $h_D = 75\,\rm Km$ and $\Delta h_D = 30\,\rm Km$ are the mean length and width of the D-layer respectively \cite{Vedantham:2013mea}. In addition to absorption, the D-layer can emit radio waves, which can add extra thermal noise. The temperature due to this additional noise can be expressed as \cite{Wang:2024lge}

\begin{equation}
    T_{\rm emit} (\nu, \theta) = T_e\left[1-\mathcal{L}(\nu, \theta)\right]. 
\end{equation}

\begin{figure}
	
	\includegraphics[width=\columnwidth]{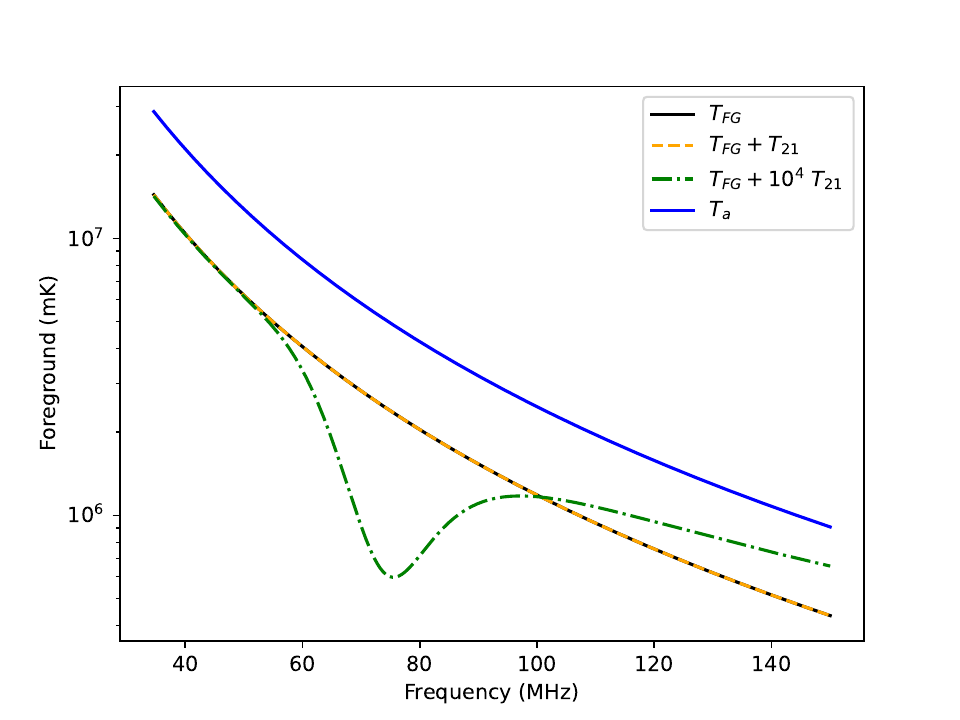}
    \caption{The black solid and orange dashed lines represent foreground signal (Eq. \ref{eq:T_FG}) in the absence and presence of $T_{21}$ signal. We rescaled the $T_{21}$ signal by multiplying it by a factor of $10^4$ to show its existence in the $T_{FG}$, shown in the green dash-dotted line. The blue solid line represents the antenna temperature $(T_a)$ in the presence of ionospheric distortion (Eq. \ref{eq:T_antenna}).}
    \label{Fitted Foreground Plot}
\end{figure}

Combining the refraction, absorption, and emission due to a static ionosphere, we can now define the antenna temperature as

\begin{alignat}{2}
    T_a = \int_{0}^{2\pi} d\phi \int_{0}^{\pi/2} & \mathcal{B}(\nu,\theta,\phi) \Big[ T_{\rm emit} + \mathcal{L}(\nu,\theta,\phi) \times  \nonumber \\ & T_{\rm sky}(\nu,\theta,\phi) \Big]\sin\theta\,d\theta,
    \label{eq:T_antenna}
\end{alignat}
where $\mathcal{B}(\nu,\theta,\phi) = \cos^2\theta+\delta\theta(\nu,\theta)\sin(2\theta)$ represents the effective antenna beam \cite{Vedantham:2013mea}. The term $T_{\rm sky}$ in the above equation represents the sum of $T_{FG}$ and $T_{21}$ signal defined in Eqs. \eqref{eq:T_FG} and \eqref{deltaTb}, respectively. In this work, we have considered $\mathcal{B}$, $\mathcal{L}$, and $T_{FG}$ independent of the azimuthal angle $\phi$. We then solve Eq. \eqref{eq:T_antenna} and plot it in Fig. \eqref{Fitted Foreground Plot}. In this figure, the black solid line represents the foreground signal $(T_{FG})$ obtained by solving Eq. \eqref{eq:T_FG}. We then add $T_{21}$ signal with the parameters $\left\{\rm T_{\rm vir}, f_{*}, f_Xf_{Xh}, f_{\alpha},\zeta_{\rm erb}\right\}$ equals to $\left\{10^4\,\rm K, 0.1, 0.2, 1, 10^{-4}\right\}$ in $T_{FG}$--- depicted in orange dashed line. Furthermore, in the green dash-dotted line, we presented the existence of $T_{21}$ signal in $T_{FG}$ by scaling $T_{21}$ to $10^4$ order. Finally, we calculated the antenna temperature due to $T_{FG}$ and $T_{21}$ signal in the presence of ionospheric distortion--- depicted in the blue solid line.

The recent detection of the global 21-cm signal from cosmic dawn by EDGES has been reported \cite{Bowman:2018yin}. In that work, the authors considered the foreground signal to be a $\log$-polynomial between the temperature and frequency as $\ln(\rm T)-\ln(\nu)$. Furthermore, they used a Bayesian technique, the Markov Chain Monte Carlo (MCMC), to extract the parameters corresponding to the foreground and global 21-cm signal. The foreground signal was represented as \cite{Bowman:2018yin}

\begin{alignat}{3}
    \frac{T_{\text{FG}}^{\rm EDGES}(\nu)}{\mathrm{K}} &= b_0\left(\frac{\nu}{\nu_0}\right)^{-2.5 +b_1+b_2\ln(\nu/\nu_0)} && e^{-b_3(\nu/\nu_0)^{-2}} \nonumber \\
    & && + b_4\left(\frac{\nu}{\nu_0}\right)^{-2},
    \label{actual Foregroud eqn}
\end{alignat}

where, $b_0$ represents an overall foreground scale factor, $b_1$ considers correction to the foreground with specified spectral index, $b_2$ considers the higher-order spectral terms, $b_3$ for ionospheric absorption effect, and $b_4$ for emission from the hot electron in the ionosphere \cite{Bowman:2018yin}. In this work, we use the linearized form of Eq. \eqref{actual Foregroud eqn},

\begin{alignat}{2}
    \frac{T_{\text{FG}}^{\rm EDGES}(\nu)}{\mathrm{K}} &\approx \left(\frac{\nu}{\nu_0}\right)^{-2.5}\Big[b_0 + b_1\ln(\nu/\nu_0) + b_2\{\ln(\nu/\nu_0)\}^2 \nonumber \\
    &\quad + b_3(\nu/\nu_0)^{-2} + b_4(\nu/\nu_0)^{0.5}\Big] 
    \label{Foreground eqn}
\end{alignat}
 to estimate these free parameters and extract the foreground and 21-cm signal from the antenna temperature using MCMC technique. In the following sections, we introduce a machine-learning technique where we will train an artificial neural network (ANN) to extract a global 21-cm signal from the total signal (i.e., from $T_a$ shown in Eq. \ref{eq:T_antenna}). We then compare our findings with the foreground and 21-cm signal extracted from $T_a$ using MCMC. We note that the total signal includes diffuse foreground $(T_{FG})$, $T_{21}$ signal, and ionospheric distortion which obscures the true nature for the foreground signal. This makes sure that we extract 21-cm signal without the explicit knowledge of foreground signal's nature. In the next section, we discuss the extraction using MCMC.

\section{Parameter inference by MCMC}\label{MCMC_section}

This section explains the inference procedure we use by deploying MCMC technique in extracting 21-cm signal. We follow a Bayesian procedure by constructing a $\log$-likelihood function, given as 

\begin{equation}
    \log[\beta(T_{a}|\theta)] = -\frac{1}{2}\sum_{i = 0}^{1024}\left[\frac{T_a^i - \rm model^i(\theta)}{T_{a, \rm error}^i}\right]^2.
\end{equation}

Here, the $\rm model (\theta)$ represents the sum of $T_{FG}^{\rm EDGES}(\nu)$ and $T_{21}$ signals from Eqs. \eqref{Foreground eqn} and \eqref{deltaTb}, respectively. The $T_a$ represents the antenna temperature obtained in the presence of $T_{FG}$ signal, $T_{21}$ signal, and ionospheric distortion by solving Eq. \eqref{eq:T_antenna}. To include an error bar in the obtained $T_a$, we added a constant uncertainty of 10 mK for all frequency bins and defined them as $T_{a, \rm error}$. The parameters $(\theta)$ that were extracted are $\left\{b_0, b_1, b_2, b_3, b_4, \nu_c, \rm T_{\rm vir}, f_{*}, f_{\alpha}, f_Xf_{Xh}\right\}$.

Before conducting the MCMC analysis, we first define the upper and lower bounds of the priors for the foreground signal $(T_{FG}^{\rm EDGES})$ and $T_{21}$ parameters, as listed in Table \eqref{tab:params}. We then initialize 50 independent Markov chains (nwalkers) for each parameter, with their starting positions specified in Table \eqref{tab:params}. To ensure an unbiased MCMC run, we normalize all parameters to the same numerical order.

\begin{table}[ht]
    \centering
    \renewcommand{\arraystretch}{1.2}
    \begin{tabular}{c c c}
        \hline
        Parameter & ~ initial position & Range \\ 
        \hline
        $b_0$   & $-3\times 10^2$ & \quad $[-6\times 10^2, -10^2]$ \\ 
        $b_1$   & $-3\times 10^3$ & \quad $[-6\times 10^3, -10^3]$ \\ 
        $b_2$   & $-2\times 10^2$ & \quad $[-4\times 10^2, -10^2]$ \\ 
        $b_3$   & $20$ & \quad $[0, 40]$ \\ 
        $b_4$   & $-1\times 10^3$ & \quad $[-3\times 10^3, 10^3]$ \\ 
        $\nu_c$ & $90\,\rm MHz$ & \quad $[70\,\text{MHz}, 110\,\text{MHz}]$ \\ 
        $\rm T_{vir}$ & $3\times 10^4\,\rm K$ & \quad $[10^4\,\rm K, 5\times 10^4\, \rm K]$\\
        $\rm f_{\alpha}$ & $3$ & \quad $[1,10]$\\
        $\rm f_Xf_{Xh}$ & $3\times 10^{-1}$ & \quad $[10^{-1},5\times 10^{-1}]$\\
        $\rm f_*$ & $2\times 10^{-2}$ & \quad $[10^{-2},5\times 10^{-2}]$\\
        \hline
    \end{tabular}
    \caption{The initial positions and upper and lower bounds on the $T_{FG}^{\rm EDGES}$ and $T_{21}$ parameters used as prior in the MCMC analysis.}
    \label{tab:params}
\end{table}

We implement the inference procedure using the publicly available \texttt{emcee} code for ensemble sampling \cite{Foreman-Mackey:2012any}. Each parameter undergoes $10^4$ sampling steps, after which we assess the convergence of the Markov chains. We compute the autocorrelation time $(\tau)$ to determine the number of steps required for the chains to become effectively independent of their previous states. For all parameters, $\tau$ remains close to unity, indicating independence among the chains. 

\begin{figure*}[ht]
    \centering
    \includegraphics[width=\textwidth]{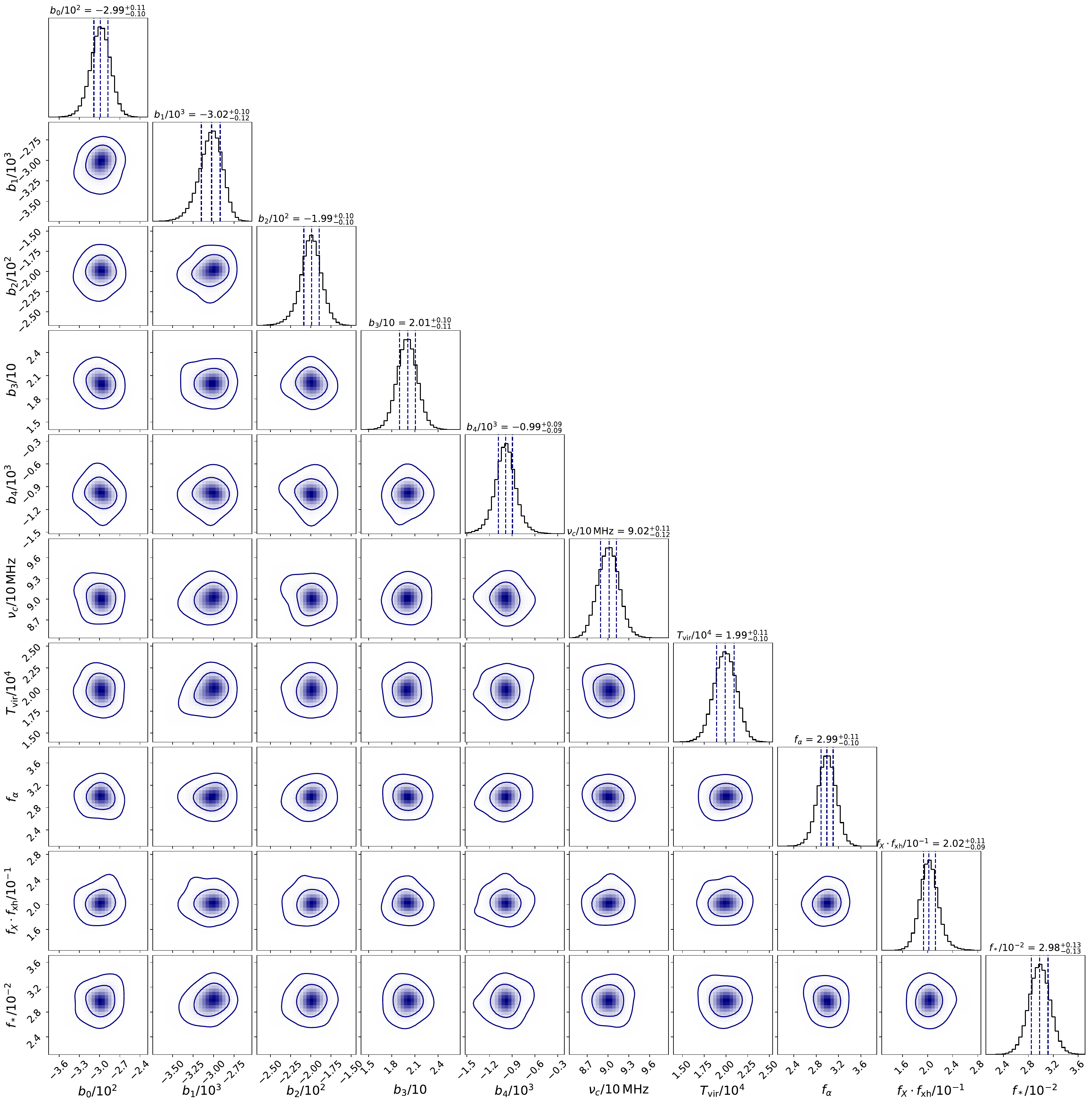}
    \caption{The one-dimensional and two-dimensional posterior probability distribution of the free parameters used to infer the antenna temperature $(T_a)$ shown in Eq. \eqref{eq:T_antenna}. The free parameters are $\left\{b_0, b_1, b_2, b_3, b_4, \nu_c, \rm T_{\rm vir}, f_{\alpha}, f_Xf_{Xh}, f_{*}\right\}$. The contour lines shows 68.3\%, and 95.5\% confidential levels corresponding to $1\sigma$ and $2\sigma$.} 
    \label{fig:MCMC}
\end{figure*}

To further verify convergence, we apply the Gelman-Rubin criterion, which ensures that the covariance of samples from an individual Markov chain matches the covariance between distinct chains at the same iteration. We quantify this using the $\hat{\cal R}$ statistic, where $\hat{\cal R}$ $<1.1$ confirms convergence \cite{1992StaSc...7..457G}. Using the \texttt{arviz} Python package\footnote{\url{https://python.arviz.org/en/stable/}}, we compute $\hat{\cal R}$ values via \texttt{arviz.rhat}\footnote{\url{https://python.arviz.org/en/stable/api/generated/arviz.rhat.html}} and find $\hat{\cal R}\approx 0.99$ for all parameters, indicating successful convergence. Finally, Fig. \eqref{fig:MCMC} presents the one-dimensional and two-dimensional posterior probability distributions of the parameters within $1\sigma$ and $2\sigma$ confidence levels.

\section{Artificial Neural Network} \label{ANN}
\subsection{Overview and construction of ANN}
This section briefly introduces the fundamental concepts of artificial neural networks (ANNs). A basic neural network consists of three principal layers: an input layer, one or more hidden layers, and an output layer. Each of these comprises of neurons; therefore, a neuron serves as the fundamental unit of an ANN. In a feed-forward and fully connected neural network, every neuron of a given layer is connected to all the neurons of the subsequent layer, and the transmission of information is unidirectional. Each of these connections is associated with a weight, a bias, and an activation function. Usually, activation functions are introduced in the layers to incorporate a non-linear behaviour. To begin the training of an ANN, a cost or error function is computed following each forward pass in the output layer. During training, the weights and biases are reassigned so that the cost function is minimal. This operation is achieved by iteratively back-propagating the errors from the final layer. A thorough explanation of the fundamental algorithm employed in a typical artificial neural network can be found in Ref. \cite{Olvera:2021jlq}.

In this work, the feed-forward network utilizes an ANN with multiple hidden layers implemented using a sequential model from Keras API and \texttt{Tensorflow}. We fixed $1024$ neurons in the input layer, which represents 1024 frequency (redshift) channels. Utilizing standard \texttt{scikitlearn} \cite{scikit-learn} and \texttt{Tensorflow} modules, we constructed our network. The number of neurons in the output layer determines the number of output parameters we aim to predict. Therefore, we fixed 4 neurons in the output layer corresponding to the parameters associated with star formation history: $\left\{\rm T_{\rm vir}, f_{*}, f_Xf_{Xh}, f_{\alpha}\right\}$ 
In the following sections, we comprehensively describe the performance and structure of the neural networks used in this work.

\subsection{Construction of training dataset}\label{training dataset}

\begin{figure}[htbp]
    \centering
    \includegraphics[width=\columnwidth]{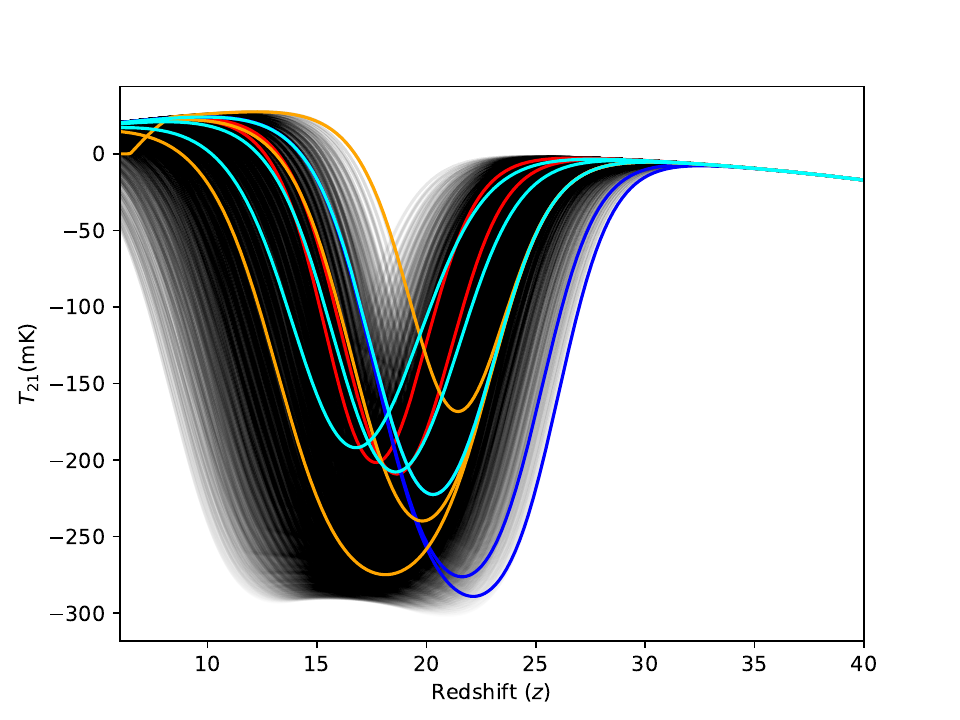}
    \caption{Construction of different 21-cm signals obtained by varying parameters $\left\{\rm T_{\rm vir}, f_{*}, f_Xf_{Xh}, f_{\alpha}\right\}$. These realisation of $T_{21}$ signals added with $T_{FG}$ and ionospheric distortion generates different $T_a$ signal, as the training dataset for the ANN.} 
    \label{21cm plot construction}
\end{figure}

In training a neural network, the quintessential quantity is the dataset. The training dataset should comprise all the data necessary for an ANN to understand the model. In this work, we train our ANN with global 21-cm signals such that it could understand the relationship between the signals and the parameters associated with star formation history described in section \eqref{Generating the global 21-cm Signal}.

We generate different realizations of 21-cm signal by varying $\left\{\rm T_{\rm vir}, f_{*}, f_Xf_{Xh}, f_{\alpha}\right\}$ from their fiducial values equals to $\left\{10^4\,\rm K, 0.1, 0.2, 1\right\}$ while fixing $\zeta_{\rm erb}$ to $10^{-3}$. We varied $\rm T_{\rm vir}$ in range $10000-30000$, $f_{*}$ as $0.01-0.1$, $f_Xf_{Xh}$ as $0.01-1$, and $f_{\alpha}$ as $1-10$. In Fig. \eqref{21cm plot construction}, we have shown the global 21-cm signals generated for this work. We then added foreground signals to this dataset to generate $T_{\rm sky}(\nu)$.
Furthermore, to include ionospheric distortion to $T_{\rm sky}$, we fixed $n_{e,f}$ and $T_e$ to $5\times 10^{11}\,\rm m^{-3}$ and $200\,\rm K$ respectively, and solve Eq. \eqref{eq:T_antenna} to generate the total signal or the antenna temperature $T_a$.

\subsection{Construction of prediction dataset}\label{Predicting data set}

\begin{figure}[htbp]
    \centering
    \includegraphics[width=\columnwidth]{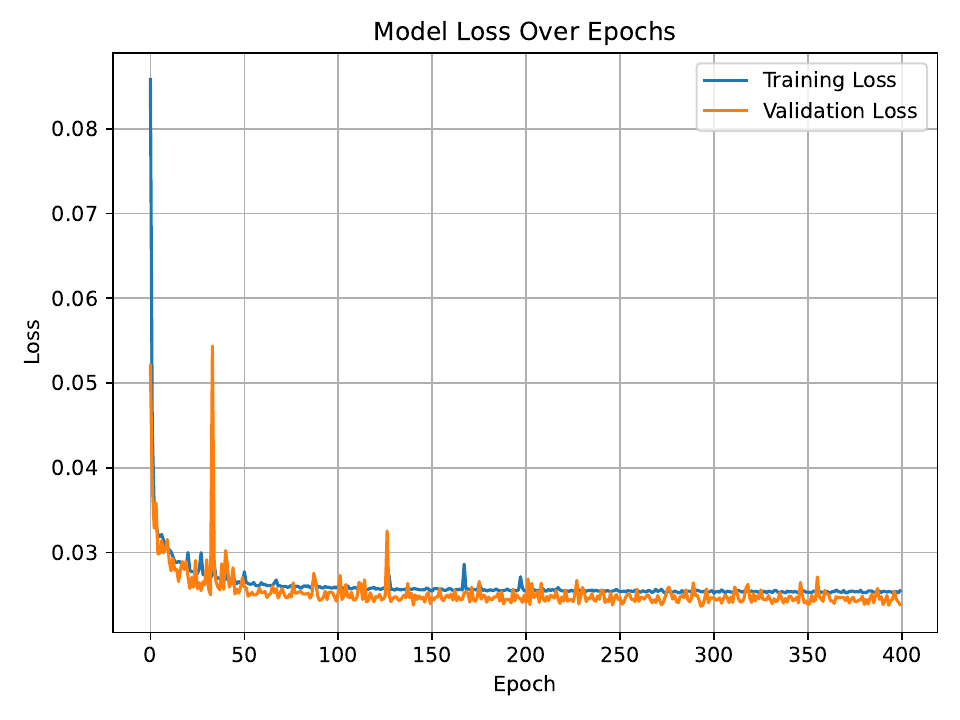}
    \caption{This graph depicts the evolution of the ANN's loss function with epochs. We can see that the test-loss function closely follows the training-loss function, as shown in the orange and blue solid lines respectively.}
    \label{fig:train_loss}
\end{figure}

\begin{figure*}[htbp]
    \centering
    \subfigure[]{

        \includegraphics[width=0.45\textwidth]{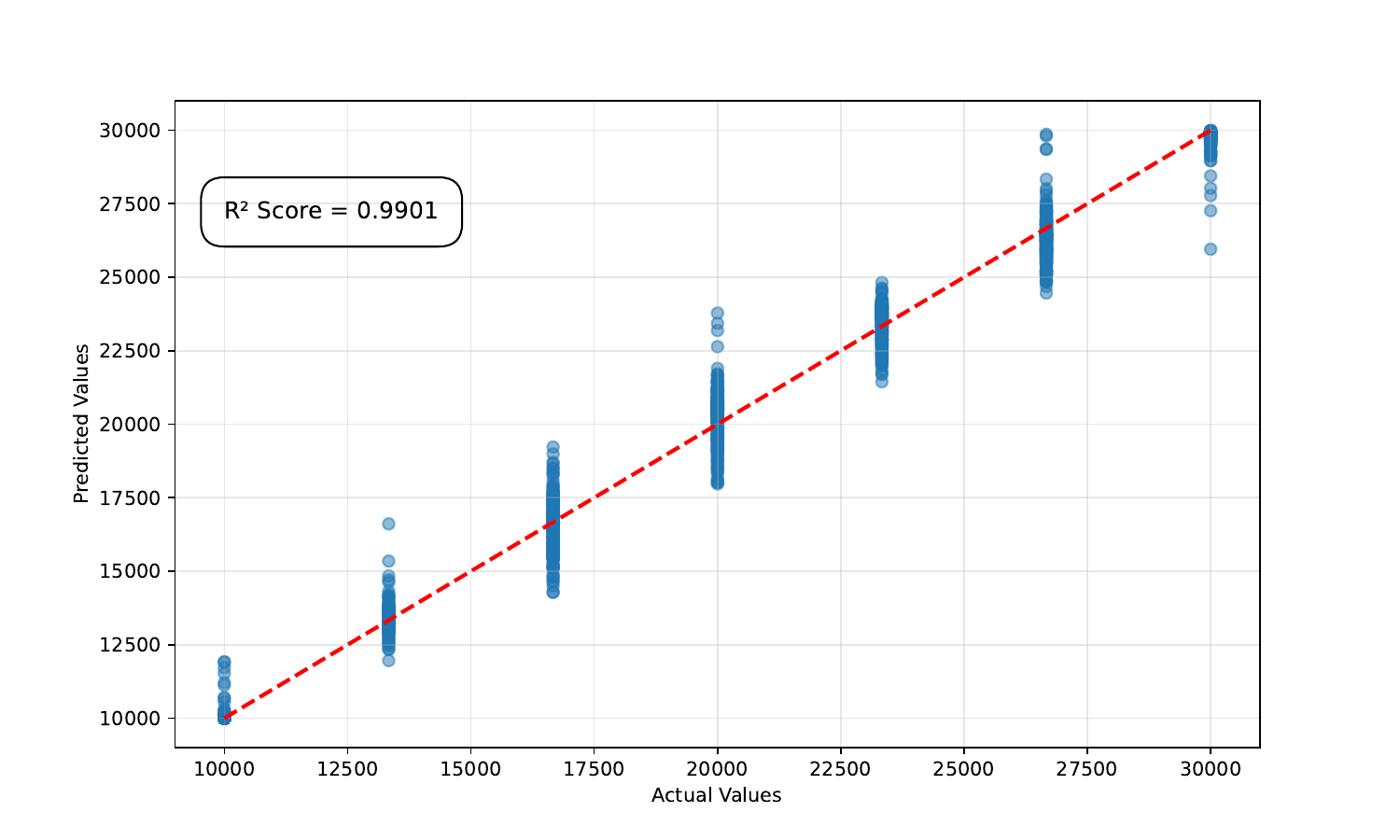}
        }
    \hfill
    \subfigure[]{

        \includegraphics[width=0.45\textwidth]{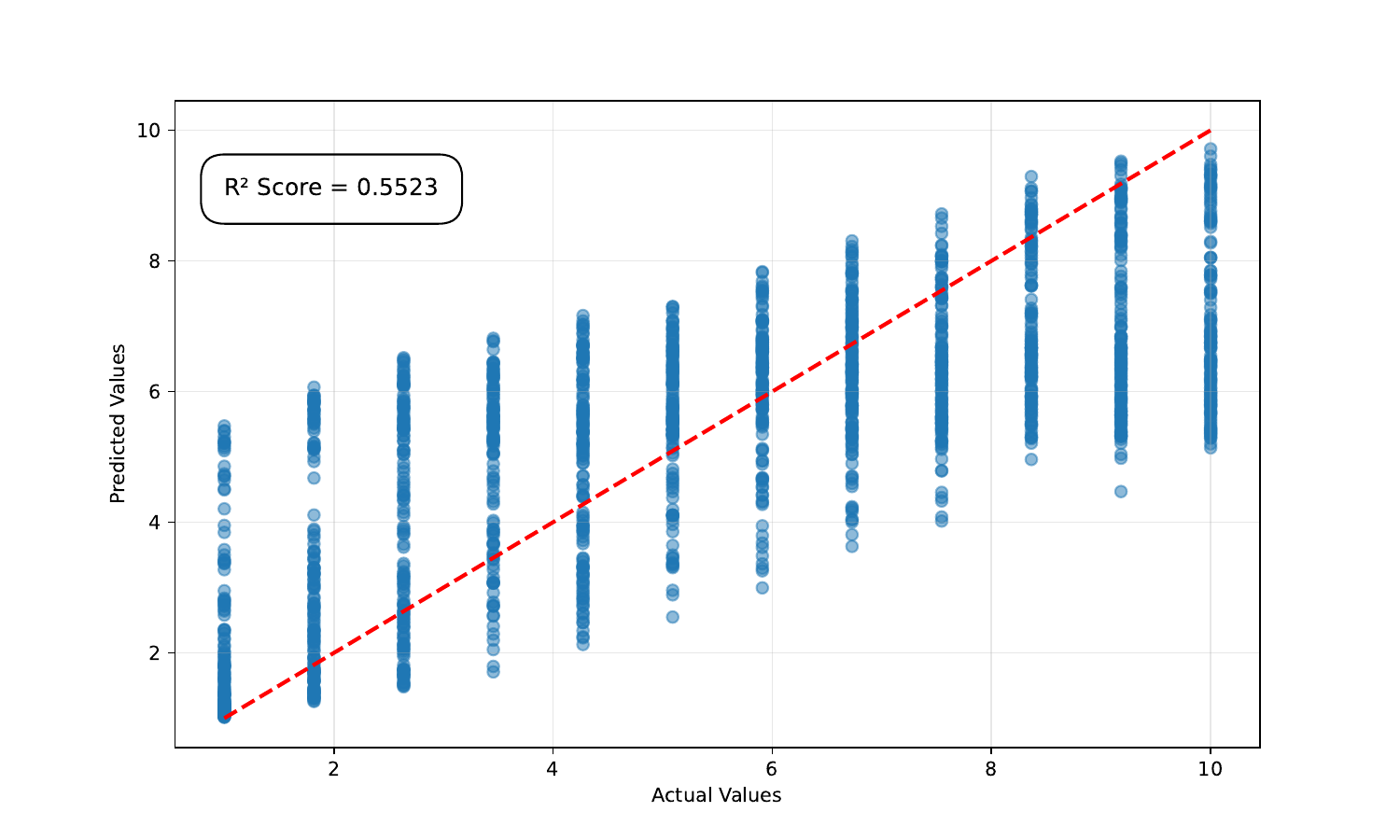}
    }
    \hfill
    \subfigure[]{
        \includegraphics[width=0.45\textwidth]{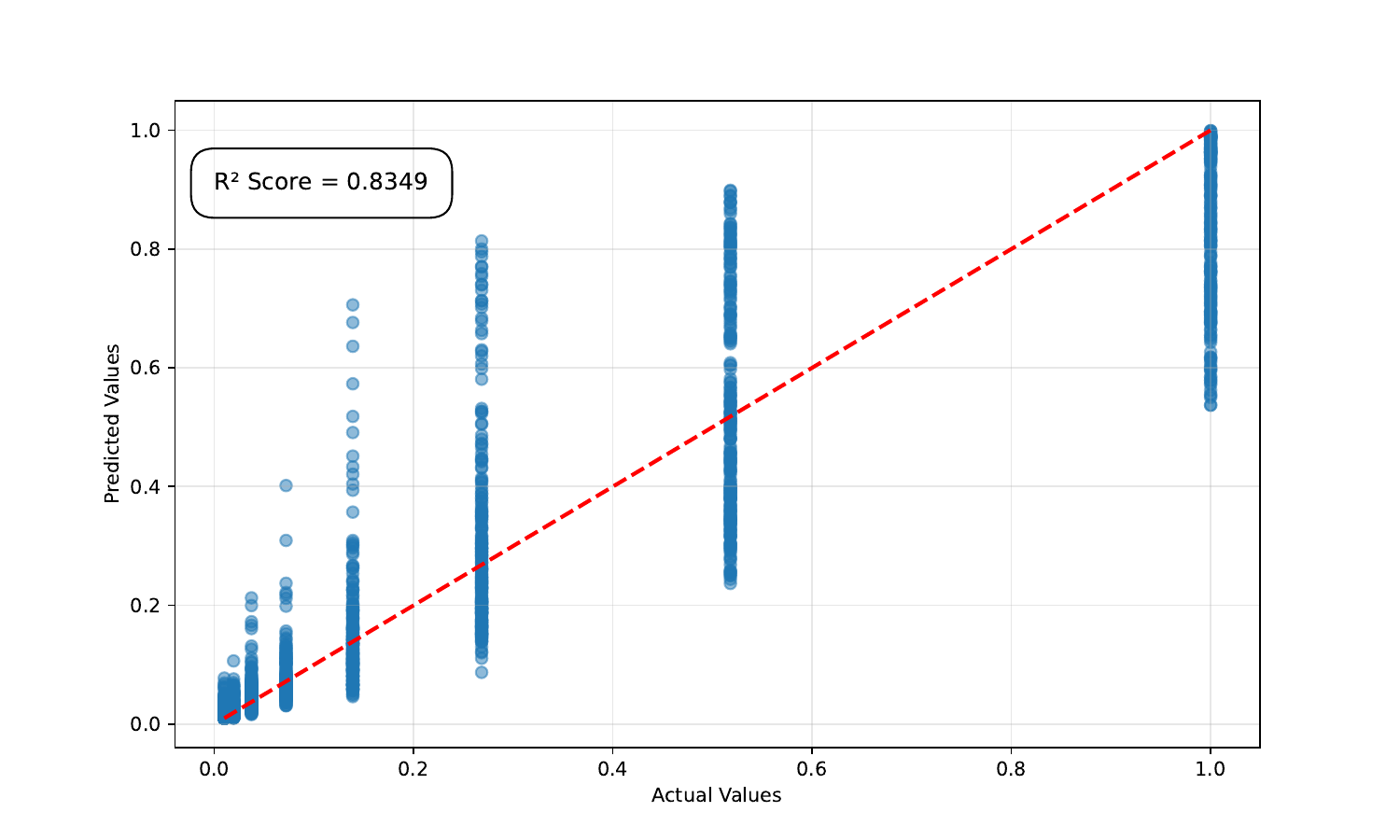}
      }
    \hfill
    \subfigure[]{
        \includegraphics[width=0.45\textwidth]{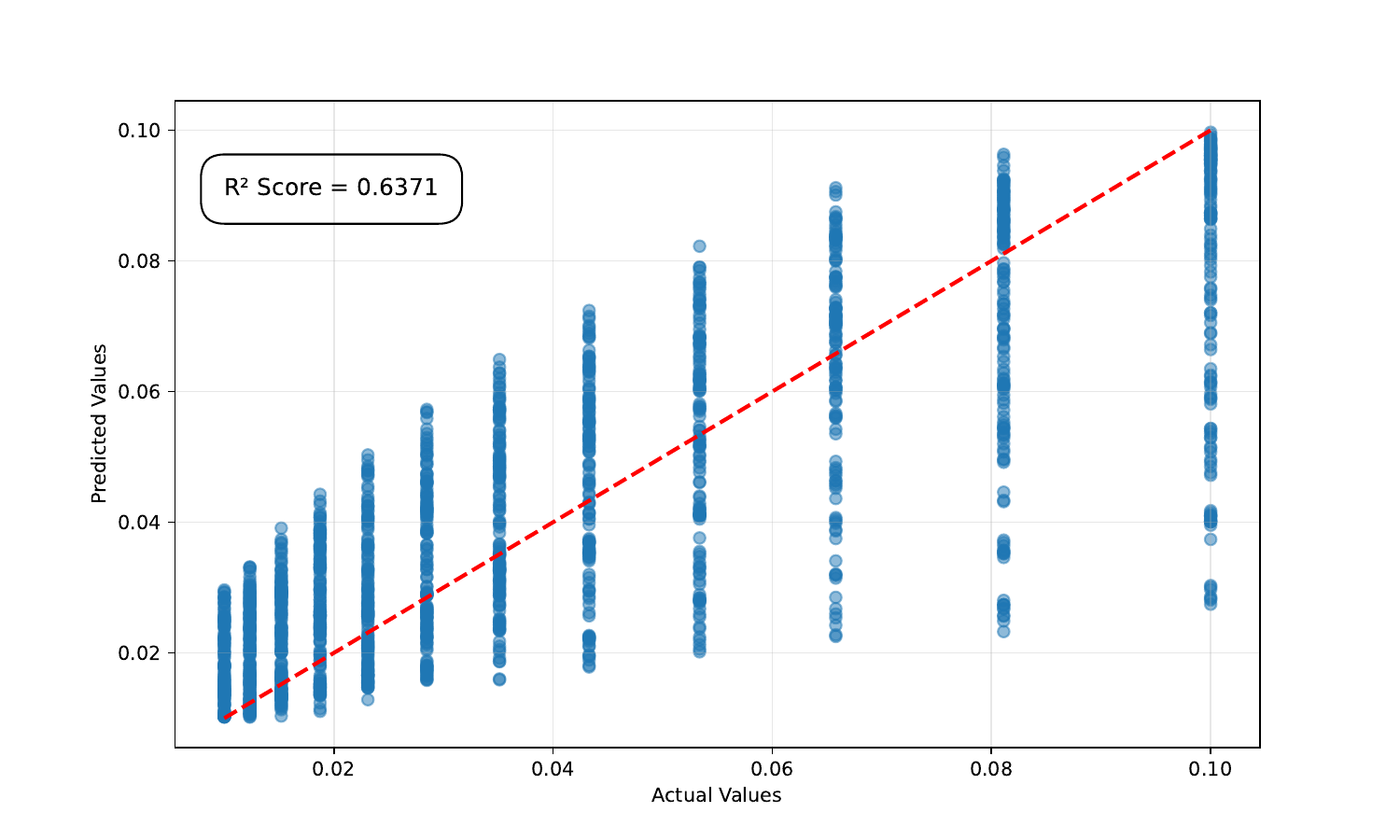}
    }
    
    \caption{The prediction of parameters $\left\{\rm T_{\rm vir}, \rm f_{*}, f_Xf_{Xh}, f_{\alpha}\right\}$. The top left and right panels represent $\rm T_{\rm vir}$ and $\rm f_*$, respectively. The bottom left and right panels represent $\rm f_Xf_{Xh}$ and $f_{\alpha}$ parameters, respectively.  
    The corresponding $\rm R^2$ scores of $\left\{\rm T_{\rm vir}, \rm f_{*}, f_Xf_{Xh}, f_{\alpha}\right\}$ are $0.9901$, $0.5523$, $0.8349$, and $0.6371$, respectively. The parameters are extracted from the trained ANN.} 
    \label{prediction}
\end{figure*}

We calculate the Root mean square error (RMSE) and $R^2$ score for each parameter estimation as a metric for all prediction datasets to depict the accuracy. The RMSE and $R^2$ score can be expressed as

\begin{equation}
    \text{R}^2 = 1 - \frac{\sum\left(X_{\text{pred}} - X_{\text{orig}}\right)^2}{\sum\left(X_{\text{orig}} - \bar{X}_{\text{orig}}\right)^2}
\end{equation}
where $\bar{X}_{\text{orig}}$ is the average of original parameters which has been summed over all training data. The $R^2$ value lies in the interval $(0,1]$ where $R^2$ $=1$ represents perfect inference.

\subsection{Training and testing of the neural network}\label{train_the_NN}

We trained the ANN using the constructed dataset mentioned earlier. To perform the training process, we first fixed two hidden layers with $64$ and $32$ neurons for optimum performance. We then implemented \texttt{elu} activation function to the input and hidden layers to introduce non-linearity to the ANN. On the contrary, we implemented \texttt{sigmoid} activation function to the output layer.  With the construction of the ANNs, we considered $8000$ realizations of $T_a$, which were split into training and testing datasets in the ratio $(8:2)$ using \texttt{sklearn} \cite{scikit-learn}. The training dataset, which is $80\%$ of the total dataset, is fed into the ANN, which was iterated over $400$ epochs. We note that the ANN does not consider the entire training data at once, instead this dataset were divided equally into small sets known as batchs. We considered the batch size to be $16$ such that each batch consists of randomly selected $(1/16)^{\rm th}$ part of the training dataset. In order to converge the training process, we introduced an error function to be evaluated after every epoch and an optimizer called \texttt{Adam} \cite{Kingma:2014vow}. The assigned error function is the Mean Squared Error (MSE) given by

\begin{equation}
    \text{MSE} = \frac{1}{\text{N}{\text{pred}}}\sum_{j=1}^{\text{N}{\text{pred}}}\left(\frac{X_{\text{orig},j} - X_{\text{pred},j}}{X_{\text{orig},j}}\right)^2.
\end{equation}

We are now all set to train the ANN and validate the training process. We first randomly selected $10\%$ of the training data to validate the training process. Therefore, after every epoch, the validation dataset is fed to the ANN to evaluate a validation MSE. The ANN is considered to be trained if the validation and training error converges together. On the contrary, if the training error decreases over epochs while the validation score becomes constant, that would suggest an under-fitting or over-fitting scenario. In Fig. \eqref{fig:train_loss}, we plot the training and validation loss over epochs in the blue and orange solid lines. We observed that the loss becomes nearly constant after $350$ epochs. Thus, we fixed the number of epochs at $400$ and saved the model for the testing process. In the next section, we discuss the $T_{21}$ parameter estimations from the ANN and MCMC, and then we will reconstruct the $T_{21}$ signals from the estimated parameters.

\section{Results and Discussions}\label{Result}

In this section, we evaluate the accuracy of the parameters predicted by the trained ANN and compare the 21-cm signal extracted using MCMC and ANN with the original signal. We begin by selecting a testing dataset that constitutes 20\% of the total dataset and remains entirely unknown to the ANN. Feeding this dataset into the ANN, we obtain parameter estimates, which we present in Fig. \eqref{prediction}. To further assess the ANN’s predictive capability, we select a random signal from the dataset--- one the ANN has not encountered before--- and estimate the corresponding $T_{21}$ signal parameters. Table \eqref{table_without_cmb} lists both the original and predicted parameter values. We find that for $\rm T_{vir}/K$, the original and predicted values are 16666.66 and 17148.66, respectively. Similarly, for $\rm f_{\alpha}, f_Xf_{Xh}$, and $\rm f_{*}$, the original and predicted values are $(5.09, 6.95)$, $(0.268, 0.333)$, and $(0.023, 0.021)$, respectively. 

\begin{figure}[htbp]
    \centering
    \includegraphics[width=\columnwidth]{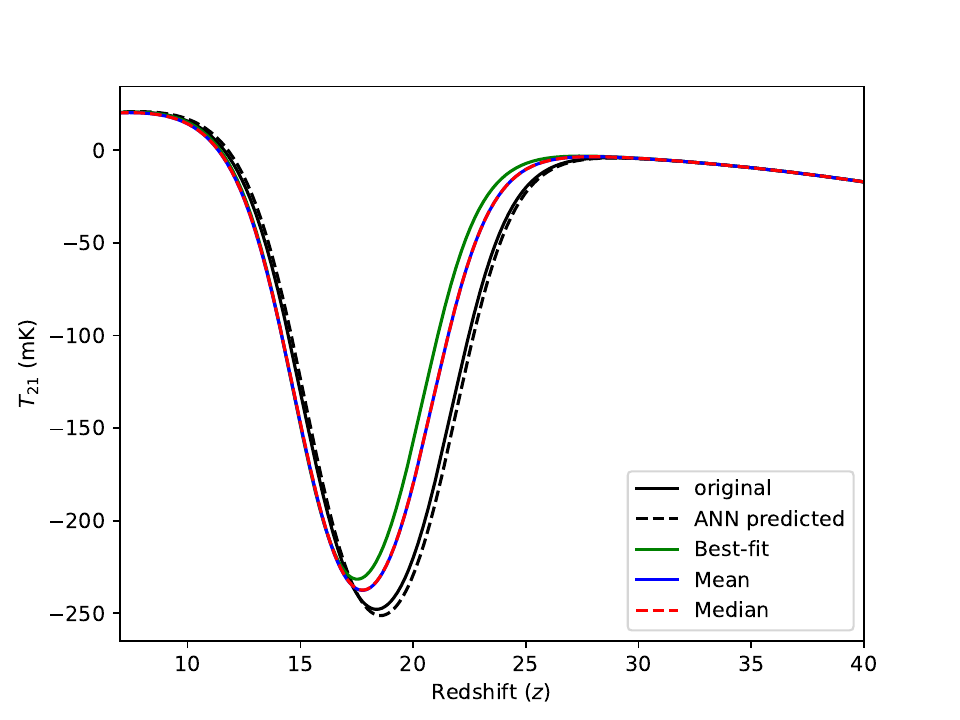}
    \caption{Reconstruction of 21-cm signal from the predicted parameters from the trained ANN and MCMC technique. The original and predicted parameters corresponding to the black solid line are presented in Table \eqref{table_without_cmb}. The black dashed line represents the reconstructed $T_{21}$ signal from the ANN predicted parameters. The green solid line represents the $T_{21}$ signal corresponding to the best-fit values of astrophysical or $T_{21}$ parameters estimated from MCMC. The blue-solid and red-dashed lines represent the $T_{21}$ signal for the mean and median values, respectively.
    }
    \label{fig:predicted_T21}
\end{figure}

We reconstruct the original and predicted $T_{21}$ signals to assess the accuracy of the extracted parameters. Fig.  \eqref{fig:predicted_T21} presents the reconstructed signals obtained from the parameters predicted by ANN and MCMC. The black solid line represents the original $T_{21}$ signal used as input for the ANN, while the black dashed line corresponds to the reconstructed signal based on the predicted parameters listed in Table \eqref{table_without_cmb}. To reconstruct the $T_{21}$ signal predicted by MCMC, we determine the best-fit values of $\rm T_{vir}, f_{\alpha}, f_Xf_{Xh}$, and $\rm f_*$ by minimizing $\chi^2$. Additionally, we compute the mean and median values of the parameter distributions for comparison. In Fig. \eqref{fig:predicted_T21}, the green solid line represents the reconstructed signal using the best-fit values, while the blue solid and red dashed lines correspond to the mean and median reconstructions, respectively.

We observe that the mean and median reconstructions closely overlap, indicating that the posterior distributions of the parameters are approximately Gaussian. In contrast, the best-fit signal deviates more significantly from the original signal, as the best-fit values are obtained by minimizing all 10 parameters simultaneously. To quantify the accuracy, we calculate the RMS residual between the original and reconstructed signals, defined as $\Delta T_{21}= T_{21}^{\rm original} - T_{21}^{\rm calculated}$. We find that the RMS residual for the ANN prediction is $\Delta T_{21}^{\rm ANN}\approx 4.57\,\rm mK$, whereas for the MCMC best-fit reconstruction, it is $\Delta T_{21}^{\text{best-fit}}\approx 16\,\rm mK$.

\begin{table}
\centering
\begin{tabular}{ccccccc}
\hline
 & $\rm T_{vir}\,\rm (K)$ & $\rm f_{\alpha}$ & $\rm f_Xf_{Xh}$ & $\rm f_*$ \\
\hline
Original  & 16666.66 & 5.09 & 0.268  & 0.023  \\
Predicted & 17148.66 & 6.95 & 0.333  & 0.021  \\
\hline
\end{tabular}
\caption{The predicted and original values of the parameters corresponding to the black solid and dashed lines in Fig. \eqref{fig:predicted_T21} by the ANN.}
\label{table_without_cmb}
\end{table}

%
%
%
\section{Summary and Conclusions}
\label{summary}
Extracting the faint global 21-cm signal buried within the strong foreground is a challenging task. As far as our understanding goes, most authors have employed Markov Chain Monte Carlo (MCMC), nested sampling or similar methods, and artificial neural networks (ANNs) for parameter space sampling. In this work, we explored both MCMC and ANN techniques for this purpose. We first modelled the $T_{21}$ signal during the cosmic dawn era in the presence of star formation and excess-radio background using ARCADE 2 detection. The star formation history was incorporated using Press-Schechter formalism. The diffuse foreground signal was modelled using the Global Sky Model, and we included additional distortions from a static and homogeneous ionosphere. Our findings indicate that ionospheric distortions can significantly increase the estimated antenna temperature in an experimental setup. Importantly, we focused on extracting the 21-cm signal without prior knowledge of the true foreground properties.

We first applied a Bayesian approach, implementing Markov Chain Monte Carlo (MCMC) to compute the posterior distribution of the foreground and $T_{21}$ signal parameters associated with the antenna temperature $(T_a)$. In this case, we modelled $T_a$ [Eq. \eqref{eq:T_antenna}] as the sum of a diffuse foreground $(T_{FG}^{\rm EDGES})$ [Eq. \eqref{Foreground eqn}] and the $T_{21}$ signal [Eq. \eqref{deltaTb}]. Notably, the polynomial fitting of the foreground radiation inherently accounts for ionospheric distortions. The resulting posterior distributions are presented in Fig. \eqref{fig:MCMC}.

Additionally, we trained an artificial neural network (ANN) to extract the $T_{21}$ signal. We first generated different realizations of the $T_{21}$ signal and combined them with a diffuse foreground component derived from the Global Sky Model, along with ionospheric distortions, to compute the corresponding antenna temperature. The ANN was then trained to learn the relationship between the input signals and the $T_{21}$ signal parameters. Our results demonstrate that a trained ANN can extract the 21-cm signal without prior knowledge of the foreground properties. The parameter predictions from the ANN are shown in Fig. \eqref{prediction}. Finally, we fed an observed $T_{a}$ signal into the trained ANN to obtain the $T_{21}$ parameters, which are tabulated in Table \eqref{table_without_cmb}. To compare the results, we reconstructed $T_{21}$ signals using both the original and predicted parameters, as shown in Fig. \eqref{fig:predicted_T21}.

We conclude that both MCMC and ANN can be used to extract the global 21-cm signal in the presence of strong foreground contamination without requiring prior knowledge of the true foreground model. MCMC is a model-driven approach that fits the data to a predefined functional form, providing posterior distributions of the free parameters along with their uncertainties. However, it is computationally expensive; for instance, on an Intel i7 $13^{\rm th}$ Gen laptop with $16$ GB RAM, running MCMC with a given number of walkers and steps takes over $15$ hours, depending on the complexity of the parameter space. On the other hand, ANN is a data-driven approach that does not require an explicit model to establish a relationship between the antenna temperature and the 21-cm signal parameters. Training an ANN with $8000$ samples takes only 1–2 minutes on the same laptop, making it computationally far more efficient than MCMC. Additionally, once trained, an ANN can be used repeatedly for rapid parameter estimation without requiring a functional form for the foreground. This makes ANN an effective tool for extracting the 21-cm signal without prior knowledge of the foreground structure. However, unlike MCMC, an ANN does not provide posterior distributions or uncertainties unless extended to a Bayesian Neural Network (BNN) \cite{9756596}. Moreover, ANN is limited to predicting parameters only within the range it has been trained on and cannot extrapolate beyond that range. Given these complementary advantages, we conclude that MCMC and ANN can be used together to improve parameter inference, with ANN serving as a computationally efficient alternative and a cross-validation tool for MCMC in future 21-cm experiments.

In future work, we intend to apply this concept and algorithm to more realistic data, like EDGES \cite{Bowman:2018yin}, SARAS 3 \cite{Singh:2021mxo}, and REACH \cite{deLeraAcedo:2022kiu}. Experiments like DARE \cite{Burns:2011wf, Burns:2017ndd}, DAPPER \cite{burns2019dark}, and FARSIDE \cite{burns2019farside} being space-based could provide less contaminated data compared to Earth-based experiments. The contamination due to RFI and ionosphere can be neglected for these experiments, thus making this work ideal for testing. Furthermore, we wish to train an ANN with datasets that include dynamical ionospheric distortion and direction dependency to make it more robust and independent of the true nature of the foreground.

\section*{Acknowledgement}
V. M would like to thank Debarun Paul for helpful discussions regarding MCMC and Anshuman Tripathi for a fruitful discussion on ionospheric distortion. P. K. N would like to acknowledge the National Centre for Radio Astrophysics--Tata Institute of Fundamental Research for providing accommodation support during the initial phase of this work. A. C. N acknowledges financial support from SERB-DST (SRG/2021/002291), India.
\section*{Code and Data Availability}
The analyses are accomplished using \texttt{Python}, broadly using publicly available modules, that is, NumPy \cite{2020NumPy-Array}, Matplotlib \cite{hunter2007matplotlib}, Pandas for data visualization \cite{mckinney2010data}, Heatmap of Seaborn \cite{Waskom2021}. The data underlying this article will be available on request to the
corresponding author.

\bibliography{main}

\end{document}